\documentclass[runningheads]{llncs}
\usepackage[T1]{fontenc}
\usepackage{graphicx}
\begin{document}
\title{From Diagnostic CT to DTI Tractography labels: Using Deep Learning for Corticospinal Tract Injury Assessment and Outcome Prediction in Intracerebral Haemorrhage}

\titlerunning{Predicting tractography from diagnostic intracerebral haemorrhage CT scans}
%
\author{Olivia N Murray\inst{1} \and
Hamied Haroon\inst{1} \and
Paul Ryu \inst{2} \and
Hiren Patel\inst{3} \and
Geroge Harston \inst{4} \and
Marieke Wermer \inst{5} \and
Wilmar Jolink \inst{6} \and
Daniel Hanley\inst{2} \and
Catharina Klijn \inst{7} \and
Ulrike Hammerbeck \inst{8} \and
Adrian Parry-Jones \inst{1} \and
Timothy Cootes \inst{1}}

%
\authorrunning{Murray et al.}
%
\institute{University of Manchester \email{olivia.murray-3@postgrad.manchester.ac.uk}\\ \and
Johns Hopkins  \and
Salford Royal NHS Foundation Trust \and
Brainomix  \and
Leiden University  \and
Isala Zwolle  \and
Radboud University Medical Centre \and
King's College London }

\maketitle              
\begin{abstract}

The preservation of the corticospinal tract (CST) is key to good motor recovery after stroke. The gold standard method of assessing the CST with imaging is diffusion tensor tractography. However, this is not available for most intracerebral haemorrhage (ICH) patients. Non-contrast CT scans are routinely available in most ICH diagnostic pipelines, but delineating white matter from a CT scan is challenging. We utilise nnU-Net, trained on paired diagnostic CT scans and high-directional diffusion tractography maps, to segment the CST from diagnostic CT scans alone, and we show our model reproduces diffusion based tractography maps of the CST with a Dice similarity coefficient of 57\%.

Surgical haematoma evacuation is sometimes performed after ICH, but published clinical trials to date show that whilst surgery reduces mortality, there is no evidence of improved functional recovery. Restricting surgery to patients with an intact CST may reveal a subset of patients for whom haematoma evacuation improves functional outcome. We investigated the clinical utility of our model in the MISTIE III clinical trial dataset. We found that our model's CST integrity measure significantly predicted outcome after ICH in the acute and chronic time frames, therefore providing a prognostic marker for patients to whom advanced diffusion tensor imaging is unavailable. This will allow for future probing of subgroups who may benefit from surgery.

\end{abstract}
\section{Introduction}
Intracerebral haemorrhage (ICH) accounts for a third of all strokes worldwide, but is responsible for a disproportionately high percentage of deaths, and ICH survivors often have severe disability. The Global Burden of Disease study showed that, between 1990 and 2010, 80\% of ICH incidents were in low and middle income countries, where access to specialist stroke centres is limited \cite{epi}. 
ICH is caused by the rupture of a blood vessel in the brain, resulting in a large body of coagulating blood (haematoma) in the brain. This causes injury by mass effect; the compression and distortion of surrounding tissues due to increased pressure in the skull. 
The toxic byproducts of the breakdown of the haematoma can cause secondary injury to the surrounding tissue and swelling around the haematoma. 

When a patient presents with stroke symptoms, they routinely receive a CT scan to identify the cause. If the stroke is caused by a clot, the patient can be treated immediately with clot dissolving medication or clot retrieval. If a bleed is identified, the rush is halted, as the only option for most patients is conservative management. Often, this diagnostic CT scan is the only form of imaging ICH patients will receive. 
There have been clinical trials of surgical haematoma evacuation, but whilst these trials have shown a modest reduction in mortality, no published studies have shown an improvement to functional outcome after ICH \cite{mistie}\cite{stitch}. 

The corticospinal tract (CST) is a white matter tract that descends from the motor cortex through the midbrain and the brain stem, and is essential for fine motor and upper limb control. Damage to the CST from stroke results in poor motor recovery \cite{cst}. CST integrity can be assessed using imaging. The gold standard imaging method for white matter is diffusion tensor imaging (DTI), an MRI technique sensitive to the diffusion of water. Water diffuses with high anisotropy along white matter fibers, and low anisotropy in grey matter. The anisotropy and direction of diffusion can be measured by applying a direction sensitive gradient pulse. To build up a diffusion tensor, this gradient must be applied in at least 6 directions. From the diffusion tensor image, we can determine white matter streamlines, and map out the white matter tracts using probabilistic tractography. 

DTI is often only used for ICH patients in a research setting, as it is not available in most centres due to cost, and is clinically challenging in acutely unwell patients.
Unlike DTI, white/grey matter contrast on a CT scan is low. For ICH patients who only receive a diagnostic CT scan, assessing the integrity of the CST from CT is challenging. 

Our aim was to develop a model that could produce tractography based labels of the CST from diagnostic CT scans alone. This tool would make white matter injury assessment available to all ICH patients with a diagnostic CT scan. This was done using paired CT and DTI scans of ICH patients to train a nnU-Net model \cite{nnU-Net} to delineate the CST on a CT scan. It was tested on a clinical trial dataset to investigate the clinical impact and utility of such a model and ask: can we use this automatic CT based assessment of CST integrity to predict motor outcome after stroke in the acute and chronic time frames?
A CST assessment tool that only requires a CT scan could be used to probe clinical trial data for a subgroup of patients with intact CSTs, who may show a improvement in functional recovery after surgery. If successful, this could be used as a selection criteria for future clinical trials.

\section{Methods}
\subsection{Data}
The data used to develop the model for this study were taken from two clinical trials. The FETCH dataset is a multi-centre dataset of 152 ICH patients, 90 of whom were available to us, and received high-directional diffusion tensor imaging, T1, and CT imaging within a 5 day time frame \cite{fetch}. The data were acquired at three different centres; Utrecht (n = 32, DTI acquired at 3T in 45 directions, Phillips), Leiden (n = 28, DTI acquired at 3T in 45 directions, Phillips) and Nijmegen (n = 32, DTI acquired at 3T in 64 directions, Siemens). 
The MISTIE III dataset was used to test the clinical utility of this model \cite{mistie}. This is a clinical trial of surgical haematoma evacuation as an intervention for ICH. It consists of 487 patients randomised to either surgical intervention or medical treatment, all of whom received a diagnostic CT scan. The demographics of the MISTIE III clinical trial are shown in Table 1.  

\subsection{Diffusion tensor image processing}
To generate high quality tractography maps, several preprocessing steps were taken. Correction for susceptibility artefacts is usually done by using a b0 image acquired in the reverse phase encoding direction. The FETCH data, however, was only acquired in one phase encoding direction. To apply susceptibility correction, PreQual software, a deep learning based pipeline,  was used to synthesise an artificial undistorted b0 image from a T1 image \cite{bedpost1}. Synthesised b0 images were generated for the whole dataset, and PreQual was used to apply FSL's TopUp and Eddy Correction tools as per FSL's FDT pipeline. 

 FSL's Bayesian Estimation of Diffusion Parameters Obtained using Sampling Techniques Incorporating Crossing Fibres (BedpostX) was implemented on the corrected DTI data \cite{bedpost2}\cite{bedpost3}. This technique is commonly used for mapping white matter tracts \textit{in vivo} from whole brain diffusion imaging. BedpostX accounts for multiple fibres in different directions passing through any voxel by using the "ball-and-stick" model of diffusion \cite{bedpost2}\cite{bedpost3}. The probability density functions of the diffusion parameters for each voxel are estimated using Markov Chain Monte Carlo sampling. A global streamline estimation can then be built by repeatedly sampling the probability density functions of the local diffusion parameters.

FSL's XTRACT tool was used to perform probabilistic tractography, and generate dissected white matter tract labels \cite{xtract1}\cite{xtract2}\cite{xtract3}. XTRACT is a wrapper for ProbtrackX, which samples the probability density functions produced by BedpostX, and adds anatomical seeds and stopping points. To allow tractography to be performed in each patient's native space, the anatomical seeds were registered from MNI to native space using EPI reg. The standard seed and stopping points for the CST are the peduncles and the motor cortex. For data from Leiden centre, anatomical seeds were selected several slices above the peduncles, due to obscuring artefacts in the midbrain.

The tractography maps were manually checked, and then linearly registered to native CT space using FSL FLIRT with a Correlation Ratio cost function. 80 tractography maps were successfully produced.

\subsection{Model training}
 
We selected the self-adapting U-Net framework nnU-Net for our model architecture \cite{nnU-Net}. This is a framework that adapts the commonly used U-Net model according to the profile of the training data, and shows a vanilla U-Net can still perform at state of the art, if the parameters are carefully selected \cite{nnU-Net2} . 

A 3D full-resolution architecture was selected, consisting of a U-Net encoder-decoder structure. The encoder comprises six blocks, including the bottleneck, and the decoder five blocks. Each block includes two convolutional layers, each with a kernel size of 3×3×3. The model was trained for 1000 epochs using a combination Dice and Cross Entropy loss on an NVIDIA a100 GPU, with a training dataset of 80 paired CT scans and CST labels (65 training, 15 testing). The data used for training and validation were selected using 5 fold cross-validation. The trained model was tested on the unseen testing dataset, and the testing ground truth and model predictions were compared using a Dice similarity coefficient. 

To assess CST interaction with the haematoma, haematoma segmentations were manually drawn for 20 diagnostic CT scans from the MISTIE III dataset. A haematoma segmentation model was trained with these paired scans and labels (16 training, 4 testing) using the procedure outlined above. 

\subsection{Clinical utility}
To assess the clinical utility of such a model, we used the MISTIE III clinical trial dataset. Inference was run over 487 diagnostic ICH CT scans to generate predicted tractography masks of the CST and haematoma segmentations for each patient. From these CST and haematoma segmentations, measures of tract integrity were calculated. Two binary metrics of tract integrity were derived from the model predictions; haematoma overlap, and tract dissection. Haematoma overlap was defined to be true if the tractography segmentation overlapped with the haematoma segmentation in any voxel. Tract dissection was defined to be true if either side of the CST was not detected in any axial slice, implying an interruption or `split' to the tract. 

The MISTIE III dataset contains various stroke outcome data at different time points. In this work, we used the National Institute of Health Stroke Scale (NIHSS) score and the modified Rankin Scale (mRS) scale. The NIHSS score consists of a series of evaluations of a patient's neurological function, including consciousness, vision, motor function, coordination, and sensation. The score ranges from 0 to 42, with higher scores indicating more severe impairment. We used a sum of all scores in the motor domain, namely facial palsy, upper limbs (left and right), and lower limbs (left and right), to generate an overall 'motor' NIHSS score, ranging from 0-19. The mRS (modified Rankin Scale) is a measure used to assess the degree of disability or dependence in daily activities among patients who have suffered from a stroke or other neurological conditions. The scale ranges from 0-6, with 0 being no impairment, and 6 being death.

To investigate outcome in both the acute and chronic setting, we selected motor NIHSS at baseline (day 1), motor NIHSS at 180 days after stroke, and mRS at 365 days after stroke to be our outcome variables of interest. 

Multiple linear regression analysis was conducted to assess the impact of CST integrity on our chosen outcome variables. We controlled for age, sex, natural logarithm of haematoma volume, intraventricular haemorrhage volume, and randomisation to surgical or medical treatment.

\section{Results}
\begin{figure}
\centering
\includegraphics[width=0.6\linewidth]{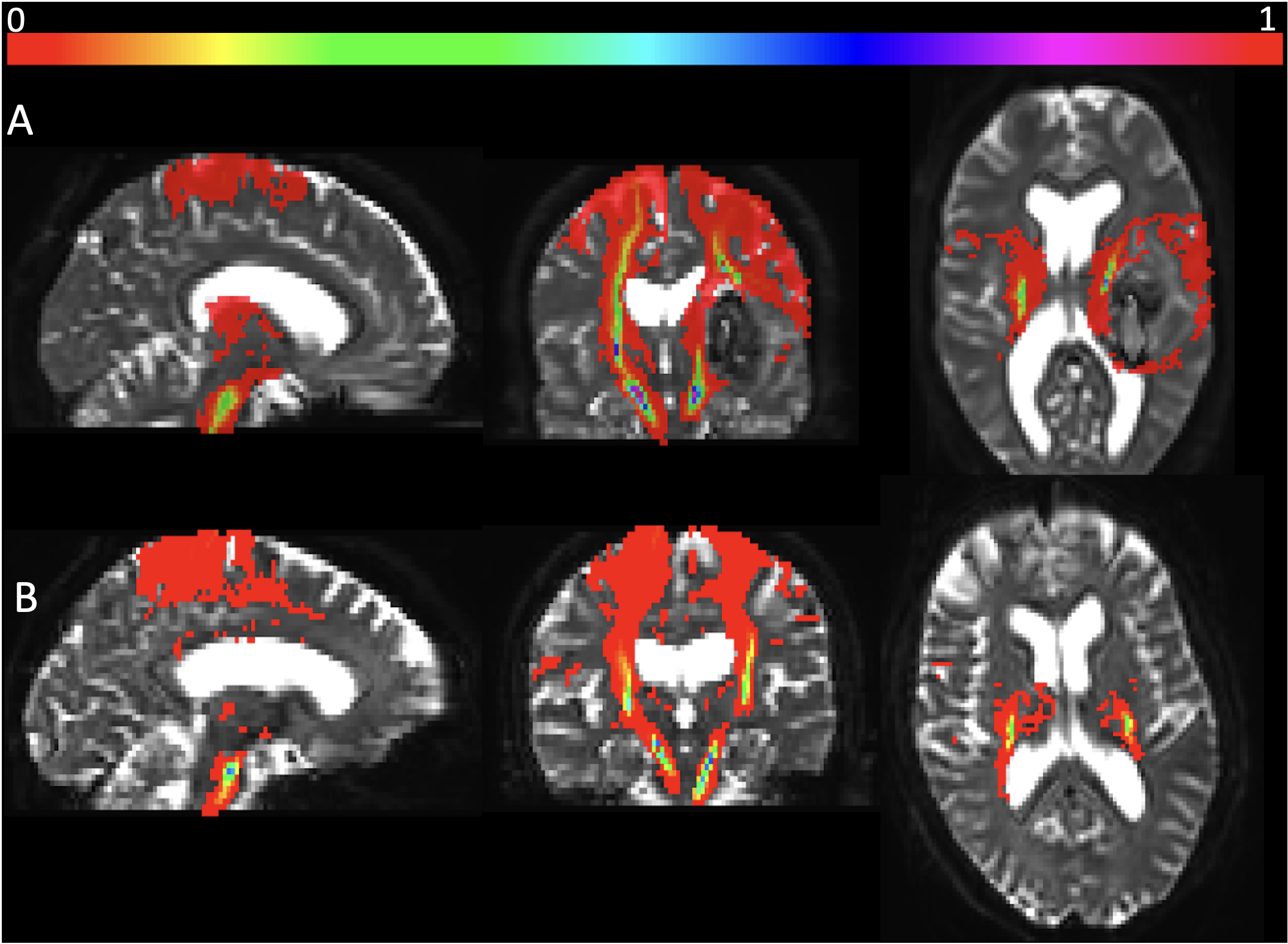}
\caption{Probabilistic tractography maps of the CST, superimposed onto the b0 volume of the DTI images for A) a patient with a haematoma involving the CST and B) a patient with a haematoma not involving the CST. }
\label{fig:fig1}
\end{figure}

Tractography maps were produced for 80 patients with DTI, T1 and CT imaging. Examples of the CST tractography maps for two patients can be seen in Figure~\ref{fig:fig1}. 
Tractography failed for 10 patients.

\begin{figure}
    \centering
    \includegraphics[width=0.7\linewidth]{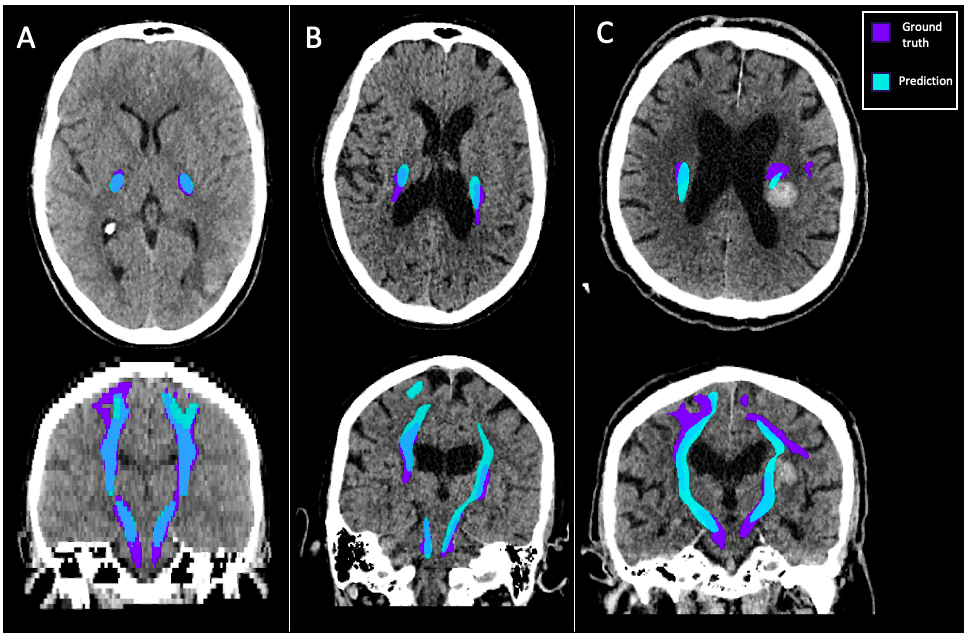}
    \caption{
Three test cases for the CST segmentation model. The ground truth DTI CST tractography is shown in purple, the model prediction is shown in cyan. A) A patient with a haematoma not involving the CST and a high DSC (DSC = 71\%) B) A patient with a haematoma not involving the CST (DSC = 52\%) C) A patient with a haematoma involving the CST (DSC = 66\%).
}
    \label{fig:fig2}
\end{figure}

 The mean DSC between the ground truth tractography segmentations in the testing dataset and the CST model's predictions was 57\%. Comparisons between the ground truth and model predictions for three patients can be seen in Figure~\ref{fig:fig2}. The mean DSC between the ground truth haematoma segmentations in the testing dataset and the haematoma model's predictions was 94\%. 

\begin{figure}
    \centering
    \includegraphics[width=0.8\linewidth]{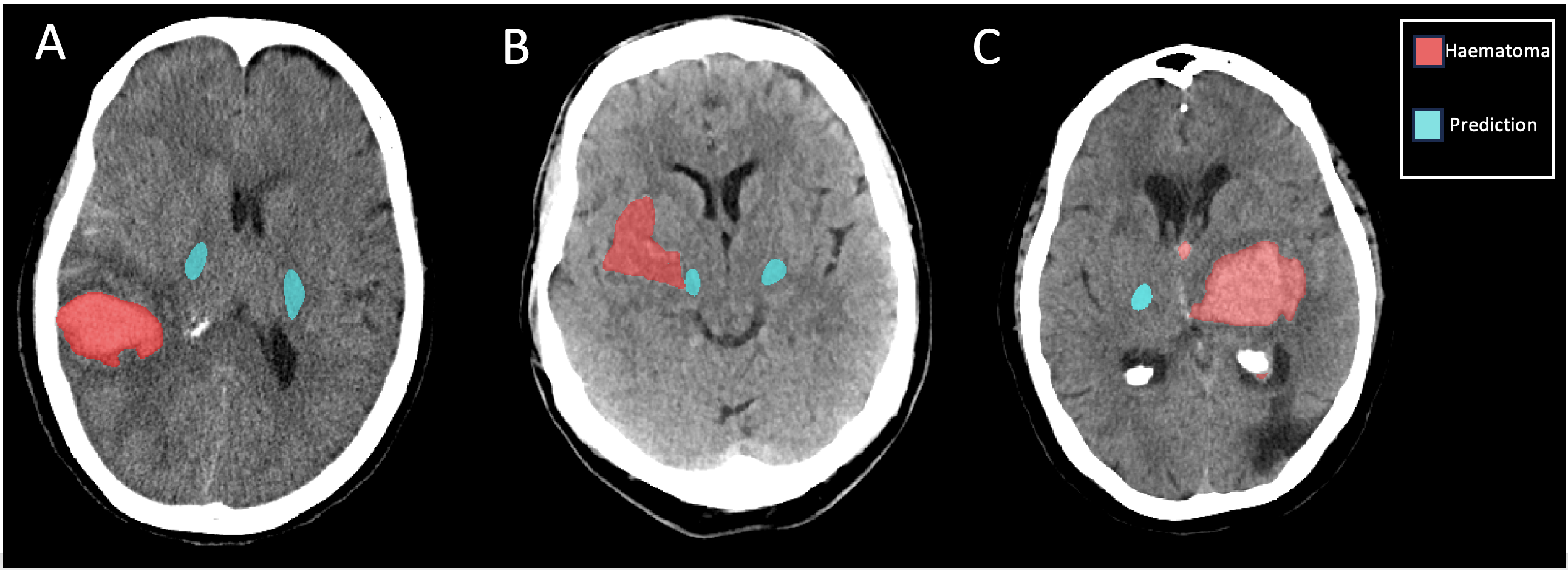}
    \caption{ Three patients from the MISTIE III dataset with predicted CST and haematoma labels. A) A patient with a haematoma not involving the CST B) A patient with CST haematoma overlap but no split tract C) A patient with no CST haematoma overlap, but a split tract.}
    \label{fig:fig3}
\end{figure}

Inference was run over diagnostic CT scans from the MISTIE III clinical trial dataset, and predicted labels of the CST and haematoma were generated for 487 patients, three of whom are shown in Figure \ref{fig:fig3}. Demographics of the MISTIE III trial, split by CST integrity, are described in Table \ref{tab1}. 

\begin{table}
\centering
\caption{Demographics of the MISTIE III clinical trial.}\label{tab1}
\begin{tabular}{|l|l|l|l|l|l|}
\hline
  &  All & CST involvement & No CST involvement & Tract split & No split\\
\hline
 & 487 & 110 & 377 & 170& 317\\
 \textbf{Age} (mean) &  61.3& 60.1& 61.6& 60.4& 61.7\\
 \textbf{Sex} Male (n) &  299 & 81& 218& 113& 186\\
   \textbf{Treatment group} & & & & & \\
Surgery (n) & 247 & 59& 188& 89& 158\\
  \textbf{Haematoma} & & & & & \\
\textbf{volume} (mean) &  43.3& 41.8& 43.7& 51.4& 39.0\\
  \textbf{IVH} & & & & & \\
\textbf{volume} (mean) & 2.4& 2.5& 2.4& 3.8& 1.6\\
  \textbf{NIHSS} \textbf{baseline}& & & & & \\
 (median [IQR])& 10 [4]& 10 [4.5]& 10 [4]& 11 [4]& 9 [4]\\
 \textbf{NIHSS day 180}& & & & & \\
(median [IQR])&  5 [7]& 6 [5]& 4 [6]& 6 [5]& 2 [6]\\
\textbf{mRS day 365}& & & & & \\
 (median [IQR])&  4 [2]& 4 [2]& 4 [2]& 4 [2]& 3 [3]\\
\hline
\end{tabular}
\end{table}

Multiple linear regression analysis investigated the effect of haematoma overlap and tract dissection on the NIHSS stroke outcome metric. Both haematoma overlap and tract dissection were significantly associated with worse NIHSS score at baseline (p = 0.028, p < 0.0001 respectively) and at 180 days post stroke (p = 0.006, p < 0.0001 respectively). Tract dissection was associated with mRS at day 365 post stroke (p < 0.0001). Other significant predictors of outcome in the model were age, treatment group and natural logarithm of the haematoma volume. The results of the multiple linear regression analysis are shown in Table \ref{tab2}.

\begin{table}
\centering
\caption{Multiple linear regression results. }\label{tab2}
\begin{tabular}{|l|l|l|l|} \hline 

  \multicolumn{2}{|c|}{}& Haematoma overlap & Tract splitting \\ \hline 

NIHSS day 1 &  $\beta$ Coefficient [95\% CI] & 0.94[0.10, 1.77]& 2.13 [1.38, 2.87]\\ \hline 
 &  p-value & 0.028*& <0.0001 ****  \\ \hline 
NIHSS day 180 &  $\beta$ Coefficient [95\% CI] & 1.32 [0.38, 2.25]& 3.55 [2.75, 4.36]\\ \hline 
 &  p-value & 0.006 **& <0.0001 ****   \\ \hline 
mRS day 365 &  $\beta$ Coefficient [95\% CI] & 0.27 [-0.05, 0.58]& 0.86 [0.58, 1.14]\\ \hline 
 &p-value& 0.097 (ns)& <0.0001 ****  \\ \hline

\end{tabular}
\end{table}

\section{Discussion}
DTI is the gold standard for white matter mapping, however, there are limitations to tractography in ICH patients. Fibre tracking can fail in close proximity to the haematoma and oedema, which are areas of low anisotropy. The mass effect of the haematoma can distort the surrounding anatomy, potentially rendering the seed points used in probabilistic tractography inaccurate. We see this in the 10 patients where tractography failed to produce reasonable white matter tracts. We also see this in the data from the Leiden centre, in which the tractography seed points had to be moved superiorly from the standard seed point in the peduncles, to avoid artefacts present in the midbrain.
However, in the 80 cases where tractography did not fail, reasonable maps of the CST were produced.  

A DSC of 57\% in the testing dataset is promising, as we are predicting the gold standard white matter imaging labels from a modality where the white/grey matter contrast is very low. A qualitative comparison of the models predictions in Figure \ref{fig:fig2} shows an underestimation of the tract volume around the edges - this could be because the edges of the tract, where the fibre density is lower, are harder to see on a CT scan. The model is able to localise the CST in cases where there are gross changes to the brain structure, such as midline shift, suggesting it has not just learnt the relative distance from the tract to the skull.

We observe `splits', interruptions in the z-axis, in the tract where the network cannot detect white matter. It is hard to say whether a split in the predicted tract is a reflection of damaged or undetectable white matter, or a failing of the model close to the haematoma. However, the fact that the presence of a `split' very significantly predicts outcome after stroke shows that this does give us valuable information about CST integrity. It could be that combining discontinuity in the tract segmentation, and overlap with the haematoma would give us a comprehensive overview of whether a patient has damage to the CST.

In spite of the above limitations of tractography in ICH patients, and the low visibility of white matter in CT scans, the statistical results of the clinical trial analysis show that we have created a useful prognostic tool. This tool could be used in any centre with a CT scan in their stroke pipeline, enabling this to be used in the developing world, where the main burden of ICH lies. This could also be used for future investigations of whether surgical intervention for ICH leads to better outcomes for patients with an intact CST.

\section{Conclusion}
DTI based tractography is the gold standard for white matter delineation. However, most ICH patients will never receive a DTI scan, as it is both clinically challenging in acutely unwell patients, and unavailable in most diagnostic pipelines. We describe a nnU-Net based model that can predict tractography labels of the corticospinal tract from CT scans alone, with a dice similarity coefficient of 57\%. We show that this model is clinically useful, and very significantly predicts outcome after stroke in the acute and chronic time frames in a large clinical trial. Our model, therefore, makes white matter injury assessment accessible to all patients who receive a diagnostic CT scan, allowing a prognosis of motor recovery from a diagnostic image for patients whom this information would be otherwise unavailable, and enabling future studies into targeted surgical interventions.

\begin{credits}
\subsubsection{\ackname} This study was funded through the MRC DTP iCASE studentship awarded to O.N. Murray

\subsubsection{\discintname} The authors have no competing interests
\end{credits}
%
%
%
%
%
%
%

\end{document}